\documentclass[a4paper,11pt]{article}
\usepackage{jinstpub} % for details on the use of the package, please see the JINST-author-manual
\usepackage{lineno}
\usepackage{xspace}
\xspaceremoveexception{-}
\makeatletter
\DeclareRobustCommand\onedot{\futurelet\@let@token\@onedot}
\def\@onedot{\ifx\@let@token.\else.\null\fi\xspace}

\def\eg{\emph{e.g}\onedot} 
\def\ie{\emph{i.e}\onedot}

\def\wrt{w.r.t\onedot} 

\makeatother

\newcommand{\sm}{SM\xspace}
\newcommand{\np}{NP\xspace}

\newcommand{\sctf}{SCTF\xspace}
\newcommand{\stcf}{STCF\xspace}

\newcommand{\ee}{\ensuremath{e^{+}e^{-}}}

\newcommand{\cpv}{\ensuremath{CP} violation\xspace}

\newcommand{\mev}{\ensuremath{\mbox{MeV}}\xspace}

\newcommand{\gevc}{\ensuremath{\mbox{GeV}/c}\xspace}

\newcommand{\mevc}{\ensuremath{\mbox{MeV}/c}\xspace}

\newcommand{\gevcc}{\ensuremath{\mbox{GeV}/c^2}\xspace}

\newcommand{\khzmm}{\ensuremath{\mbox{\,kHz}/\mbox{mm}^2}\xspace}
\newcommand{\mhzmm}{\ensuremath{\mbox{\,MHz}/\mbox{mm}^2}\xspace}

\title{\boldmath Performance of the FARICH-based particle identification at charm superfactories using machine learning}

\author[a]{M.~Chadeeva,}
\author[a]{P.~Rogozhin}
\author[a,b]{ and T.~Uglov}
\affiliation[a]{P.~N.~Lebedev Physical Institute of the Russian Academy of Science,\\
53, Leninskiy prospekt, Moscow, Russia}
\affiliation[b]{Higher Scool of Economics, Moscow, Russia}

\emailAdd{mchadeeva@gmail.com}

\abstract{
    A detailed study of the particle identification by the Focusing Aerogel Ring Imaging CHerenkov subsystem at the future charm superfactory detector is presented. The dedicated signal ring reconstruction algorithm is implemented in the detector simulation, the algorithm performance is tested with single particles generated within the Aurora framework. Two Boosted Decision Trees-based classifiers for the particle identification have been developed for various assumptions about photosensor noise levels. The approach is validated with the analysis of the $D^0\to K\mu\nu_\mu$ decays, for which the systematic uncertainty and background contribution related to the $\pi/\mu$ separation performance can be minimised due to high efficiency of the particle identification algorithm.
}

\keywords{Charm Superfactories, Ring Imaging Cherenkov detector, particle identification, Boosted Decision Trees}

\begin{document}
\maketitle
\flushbottom

\section{Introduction}
\label{sec:intro}
The Standard Model (\sm) --- a theory describing our Universe at the most fundamental level  --- despite of its obvious successes is far from completeness. Incapability to describe the nature of 19 free parameters arising in the model, fail to incorporate dark matter and dark energy phenomena motivate searches for the effects beyond the \sm  --- a New Physics (\np). There are two main approaches for the \np  detection. Collisions at the highest available energies, \eg in the LHC experiments, are sought to yield particles of new types. Another way is to precisely measure the \sm processes and search for indirect manifestations of the \np in the interference  processes such as \cpv or rare and forbidden particle decays. The latter approach requires the huge amount of accumulated data and the outstanding detector performance. The new beam collision scheme 'crab-waist' proposed by Raimondi~\cite{ref:raimondi} in 2006 made a revolution in electron-position collider technologies opened the opportunity for creation of flavour superfactories. The future charm superfactories, a next generation symmetric \ee colliders,  and their detectors are designed  for precision measurements of the parameters of charmed hadrons and $\tau$ leptons. There are two competing projects, one Russian Super Charm-Tau Factory (\sctf)~\cite{ref:Achasov:2024} and one Chinese Super Tau-Charm Facility (STCF)~\cite{ref:Achasov2023}, sharing physical tasks and based on similar experimental techniques. 

The lepton flavour universality (LFU) tests is one of the most challenging tasks in the charm superfactories physical program. According to the \sm, weak interaction is not sensible to the lepton flavour, \ie  in similar processes weak matrix elements for all lepton flavours must be exactly the same. The  difference in decay branching ratios originates from the phase space  and hadronic form-factors.  For the semileptonic $D$ meson decays the observable to test LFU is the branching fractions ratio:
\begin{equation}
\label{eq:br_ratio}
    \mathcal{R}_{\mu/e}= \frac{\mathcal{B}(D\to K\mu\nu_\mu)}{\mathcal{B}(D\to Ke\nu_e)}.
\end{equation}
The accuracy for $\mathcal{R}_{\mu/e}$ measurement is strongly limited by detector particle identification (PID) capabilities due to inevitable backgrounds, for example from $D^0\to K^-\pi^+\pi^0$ decays, which mimic the signal $D^0\to K^- \mu^+ \nu_\mu$ in case of $\pi/\mu$ misidentification. The combined PID system for charm superfactories is aimed to provide very efficient $K/\pi$ and $\pi/\mu$ separation within all reachable momentum range. 
In this paper, performance of the dedicated PID subsystem proposed for the \sctf is studied using single particles and  semileptonic $D$ meson decays.

\section{Particle identification subsystem at charm superfactories}
\label{sec:PID}
\subsection{Detector technologies for charm superfactories}
The detector for a charm factory is a general purpose high hermiticity magnetic spectrometer optimized for the precision studies of the various particles production and decay properties. 
Tracking system consists of the inner tracker aimed to provide superior spatial resolution at interaction point (IP) and the main tracker for precise momentum measurement. 
For the inner tracker  micropattern gaseous detector, CMOS\footnote{Complementary Metal–Oxide–Semiconductor} maps  and time projection chamber options are considered. 
The technology for the main tracker is a drift chamber in the $1.0$~T axial magnetic field provided by a superconducting solenoid.
Particle identification system is based on the Cherenkov light detection: DIRC-like time-of-flight or RICH\footnote{Ring Imaging CHerenkov detector} detectors. 
Photons are detected with the electromagnetic calorimeter, e.g. build from pure-CsI crystals to cope with high backgrounds.
Muons and $K_L$'s are identified in the instrumented (resistive plate chambers or organic scintillator)  flux-return yoke.   The detailed description of the \sctf and \stcf detector designs can be found elsewhere~\cite{russianCDR,ref:Achasov2023}. Unless the opposite is  explicitly  stated, the \sctf project paradigm is assumed throughout the paper.

The Focusing-Aerogel Ring-Imaging Cherenkov (FARICH) detector is one of the options proposed for the particle identification system at the \sctf. Its main task is the separation of muons, charged pions and kaons in the momentum range from $\sim$0.5 to $\sim$1.5~GeV/c. The FARICH radiator is composed from few layers of aerogel with slightly different refraction indices resulting in focusing of Cherenkov photons in the thinner ring at the photodetector plane compared to that from a homogeneous radiator. The silicon photomultiplier matrices are considered as an option for photon detection devices. The detailed description of the concept as well as few test beam results for the prototype can be found in ref.~\cite{Barnyakov:2024ydq}.

\subsection{Simulation and digitisation of the FARICH subsystem}
\label{sec:simfarich}
\begin{figure}[h]
\begin{center}
        \includegraphics[width=.5\textwidth]{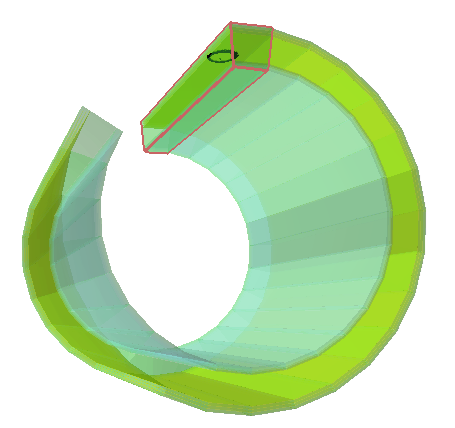}
\end{center}
    \caption{\label{fig:eventdisplay} Event display of the FARICH barrel subsystem response to single pion: aerogel layers (gray-blue) and photodetector planes (light green) grouped in modules. The hits produced by Cherenkov light are shown in black in the module highlighted with red lines. Some modules are removed for better visibility.}
\end{figure}

For charm factories detector simulations and data processing, the dedicated software packages are under development. As many other modern HEP projects, both Aurora package~\cite{Belozyorova:2021ztt} for the \sctf and OSCAR package~\cite{Huang_2023} for the \stcf are based on the DD4hep~\cite{frank_markus_2018_1464634} framework.  
Within the \sctf Geant4~\cite{ref:geant4} detector model, the FARICH system is placed between the tracker and electromagnetic calorimeter. Hereinafter the simulations are discussed of the FARICH barrel part, which has the following layout: it is composed of 27 equivalent trapezoid blocks starting at 800 mm from the IP with the total thickness of 40 mm, each block contains 4 aerogel layers (radiator), the layer of sensitive detector material is separated from the radiator by 200-mm air gap. The number of aerogel layers and their optical properties in the FARICH subdetector model are identical to those in ref.~\cite{Barnyakov:2023nng}. The schematic view of the FARICH subsystem of the \sctf detector with the ring from single simulated pion is shown in figure~\ref{fig:eventdisplay}.

The optical photon transport simulations are known to be extremely time consuming. There are two simulation modes within the Aurora package --- with and without Geant4 optical photon functionality implemented for the FARICH subdetector. In this study, the signal reconstruction and particle identification performance have been investigated using particle propagation from interaction point with 1~T magnetic field up to FARICH subdetector and subsequent full Geant4 simulation of optical photon transport through the FARICH components. The particle momentum vector components at the track end point in the tracker is taken from generator-level information and are projected to the entrance point at the FARICH radiator.

The simulated response of the FARICH subsystem is produced by Cherenkov photons generated and propagated to sensitive detectors by Geant4. To get realistic hits, the quantum and geometrical efficiency of photosensors is implemented as digitisation procedure of the photon detection: the Geant4-generated Cherenkov photon spectrum is convolved with the quantum efficiency curve and an additional factor to account for an effective geometrical efficiency. The energy dependence of quantum efficiency is taken from ref.~\cite{ref:ketek} for the KETEK 8x8 SiPM matrix (considered as one of the photosensor options) with the pixel (SiPM) size of 3.36$\times$3.36~mm$^2$ and geometry fill factor of~88\%. Afterwards, photon arrival points in the detector plane are put on a square grid with a size corresponding to the aforementioned pixel size. One or more photons registered in a single pixel force the pixel to produce one hit. The Cherenkov ring in the FARICH subdetector from a simulated pion is shown in figure~\ref{fig:eventdisplay}.

The noise hits corresponding to various assumptions about SiPM dark count rates (DCR) are also superimposed at the digitisation step. The number of noise hits for each event is sampled from Poisson distribution with the mean corresponding to the expected noise level in a pixel area. Two spatial coordinates of each noise hit at the photodetector plane are generated uniformly, the times of hits are uniformly distributed over the event time frame of 7~ns. 

\subsection{Ring reconstruction algorithm}
\label{sec:ringreco}
 The typical reconstructed hit image from a single particle in the FARICH is not a regular ring but a distorted elliptical figure because of the particle trajectory inclination with respect to the photodetector plane and the additional air gap between the radiator and the photosensors. The distortion effect depends on the particle incident angle. Moreover, the restricted efficiency of photon detection results in stochastic fluctuations in number of hits for the same particle velocity~$\beta$. 

To take these peculiarities into account, a dedicated algorithm of ring reconstruction developed and tested on simulated events in the FARICH prototype \cite{Barnyakov:2023nng} is used. The value of~$\beta$ is extracted using the incident particle momentum and direction at the at the photodetector plane, relative number of hits per ring around the projected particle momentum, and photon arrival times. The intersection of the track projection from the tracker with the FARICH photosensor plane serves as a "seed" for the Cherenkov ring reconstruction. 

\begin{figure} [h!]
\begin{center}
        \includegraphics[width=.6\textwidth]{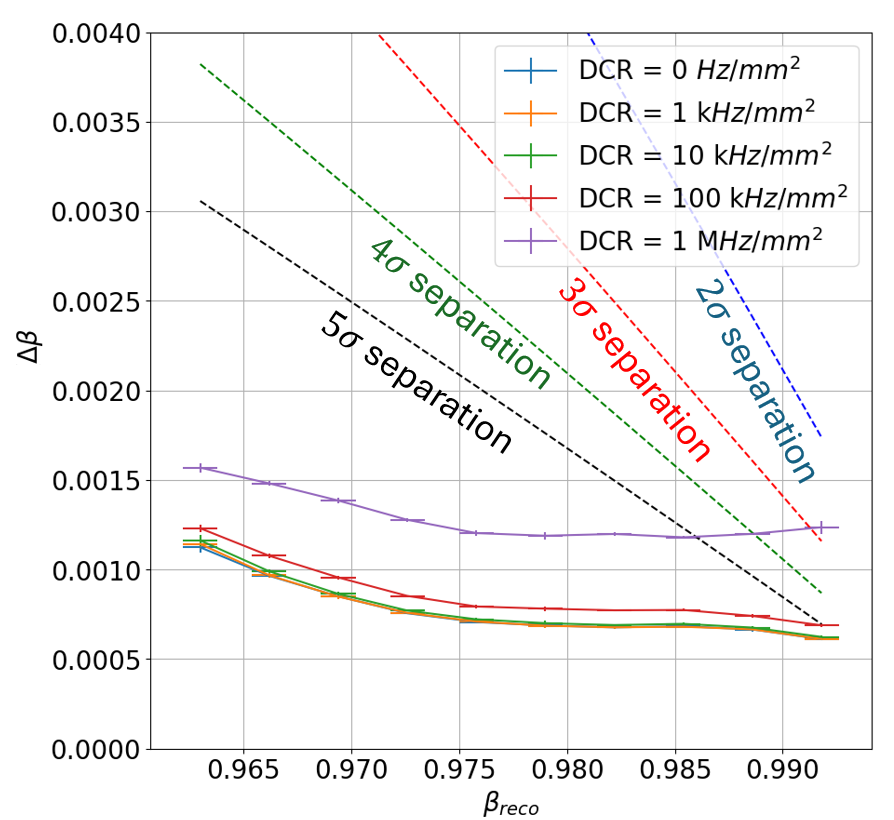}
\end{center}
    \caption{\label{fig:dbeta_vs_beta} Uncertainty of $\beta_{\mathrm{reco}}$ from the simulations of single pion response in the FARICH subdetector of the \sctf at different DCR levels. See text for details about $n\sigma$ separation.}
\end{figure}

The analysed value is the ratio $S = N(R,T)/R$, where $N$ is the number of hits calculated  in bins at the ($R, T$) plane, where $R$ is the hit radius  \wrt the "seed" point defined above and $T$ is the hit time. The outputs of the algorithm are the reconstructed radius, $R_{\mathrm{reco}}$, that corresponds to the maximum value of the ratio, $S_{\mathrm{max}}$, and the signal z-score parameter, $Z_{\mathrm{DCR}}$, that is defined as follows:
\begin{equation}
\label{eq:z_score}
Z_{\mathrm{DCR}} = \frac{S_{\mathrm{max}} - S_{\mathrm{DCR}}}{V_{\mathrm{DCR}}},
\end{equation}
\noindent where $S_{\mathrm{DCR}}$ and $V_{\mathrm{DCR}}$ are the mean and standard deviation of the ratio $S$ estimated from the pure noise events for a given DCR value, respectively. The parameter $Z_{\mathrm{DCR}}$  characterises the signal significance over the given photosensor dark count rate and is further used in the identification.
Using the reconstructed radius, a particle velocity $\beta_{\mathrm{reco}}$ can be calculated. The algorithm was demonstrated to provide the reconstruction accuracy, $\Delta\beta$, better than 0.001 for $\beta <$~0.999 at incident angles up to 45\textdegree \, and up to photodetector dark count rates of $10^5\,\mathrm{Hz}/\mathrm{mm}^2$. The performance slightly decreases at higher dark count rates expected due to damaging of SiPMs  by irradiation at the \sctf conditions. The accuracy of velocity reconstruction by the FARICH subsystem for single pions simulated in the \sctf detector is shown in figure~\ref{fig:dbeta_vs_beta} for different DCR values of photosensors. 
Assuming the same accuracy for pions and muons, the difference between means of pion and muon reconstructed mass distributions is larger than 5 standard deviations ($5\sigma$ separation) up to $\beta=0.985$ even for the worst case of $\mathrm{DCR} = 1\mhzmm$. For $\mathrm{DCR} \leq 100\khzmm$, the $5\sigma$ FARICH-based separation is available up to $\beta=0.995$.
The alternative reconstruction approach based on machine learning is presented in ref.~\cite{Shipilov:2023llf} and demonstrates comparable performance.

 The reconstruction of particle velocity is an intermediate step in the particle type identification. The accuracy of velocity estimates discussed above is shown to demonstrate the performance of Cherenkov ring reconstruction. In the next step, the obtained ring parameters can be combined with the momentum measurements from tracker system with the aim to identify a particle specie.

\subsection{Boosted Decision Trees classifier}
\label{sec:bdt}
The particle identification landscape at charm superfactory energies is expected to contain a number of species: electrons, muons, pions, kaons and protons. In the tracker with radius of 800~mm positioned inside the 1\,T magnetic field, only particles with momenta above $\sim310$\,\mevc will reach the FARICH subdetector at incident angles smaller than 45\textdegree. The tracks with larger incident angles are not considered for analysis due to the poor performance of the ring reconstruction algorithm. To produce Cherenkov light in the radiator with given parameters (see section~\ref{sec:simfarich}) the particle momentum should exceed approximately 350\,\mevc for muons, 500\,\mevc for pions and 1800\,\mevc for kaons. Electrons and protons are excluded from the analysis: electron always produces maximal Cherenkov ring, while proton does not produce any ring. 
Moreover, electrons are expected to be reconstructed in the electromagnetic calorimeter with high efficiency.
A particle specie is identified using the Boosted Decision Tree (BDT) technique.
For the current study, the $\mu / \pi / {\mathrm{K}}$ classifier is considered.  

\subsubsection{BDT training and optimisation}
Two classifiers based on the BDT approach have been constructed and trained using the XGBoost library version 2.1.2~\cite{ref:xgboost} for two noise levels. 
Four variables serve as BDT inputs: particle momentum and incident angle at the photodetector plane, reconstructed radius $R_{\mathrm{reco}}$ and signal z-score parameter $Z_{\mathrm{DCR}}$ dependent on the chosen photosensor dark count rate (see section~\ref{sec:ringreco}). A particle momentum and incident angle in the real detector will be reconstructed in the tracker and extrapolated to the radiator plane closer to the tracker. 

\begin{figure}
    \includegraphics[width=.48\textwidth]{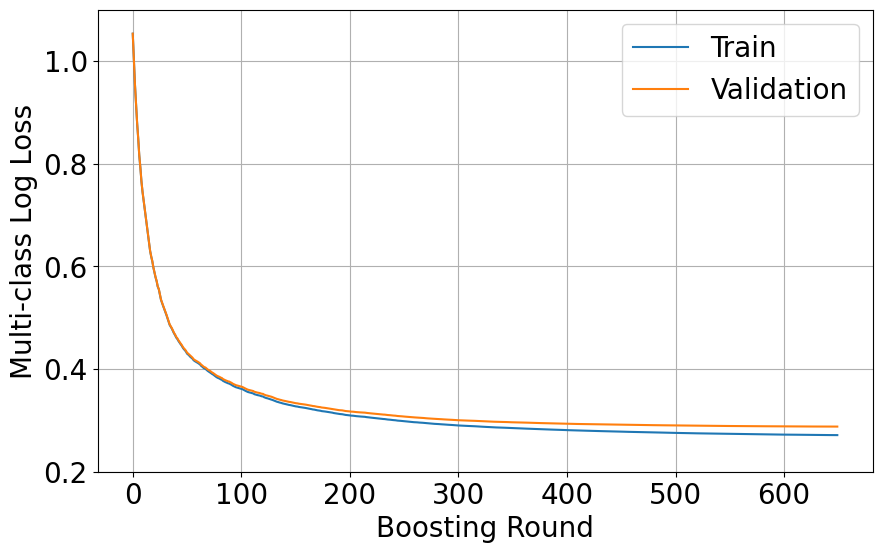}
    \includegraphics[width=.48\textwidth]{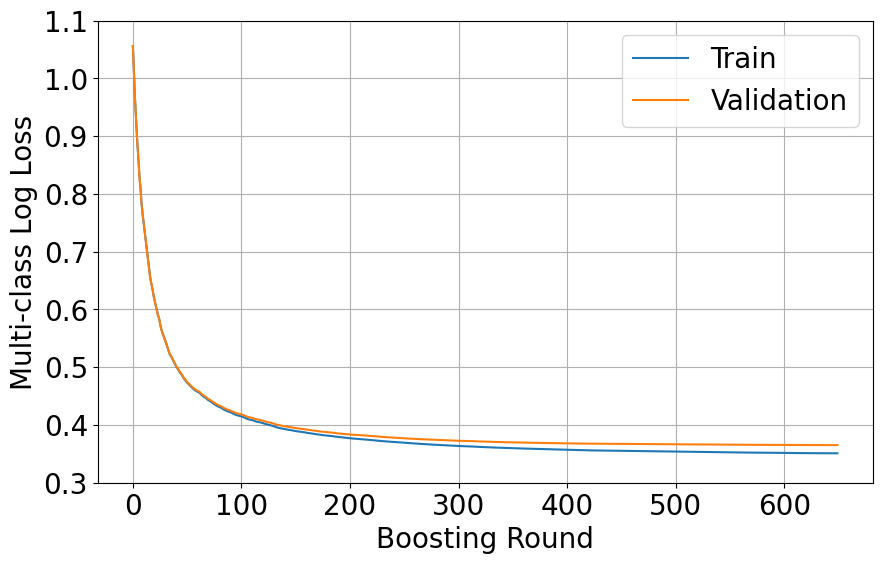}
    \caption{\label{fig:bdtloss} Multiclass logistic loss function dependence on the number of boosting rounds for training (blue) and validation (orange) samples for $\mathrm{DCR} = 100\khzmm$ (left) and $\mathrm{DCR} = 1\mhzmm$ (right).}
\end{figure}

As can be seen from figure~\ref{fig:dbeta_vs_beta}, the reconstruction algorithm shows very similar performance for dark count rates from  100\khzmm and below. For this reason, two separate classifiers have been trained for noise levels $100\khzmm$ and $1\mhzmm$. For training, 450000 events were generated, 150000 for each muon, charged pion and charged kaon samples. The particle momentum for each particle type is uniformly distributed from the minimum momentum detected in the FARICH subsystem up to 2500~$\mev/c$. Figure~\ref{fig:bdtloss} shows the loss function behaviour during the training. It can be seen that for both noise levels, about 250 boosting rounds are enough to achieve good performance without overtraining.

The classifier hyperparameteres were optimized on the grid by maximizing the performance characteristic ROC-AUC  using the tool provided in the XGBoost library. 
Additional 50000 events per each particle specie (150000 events in total) were generated for test samples, which are used to demonstrate the classifier quality estimates. Both training and test samples have the same equal fractions of three particle species.

\subsubsection{BDT performance for single particles}
\begin{figure}[h]
    \includegraphics[width=.49\textwidth]{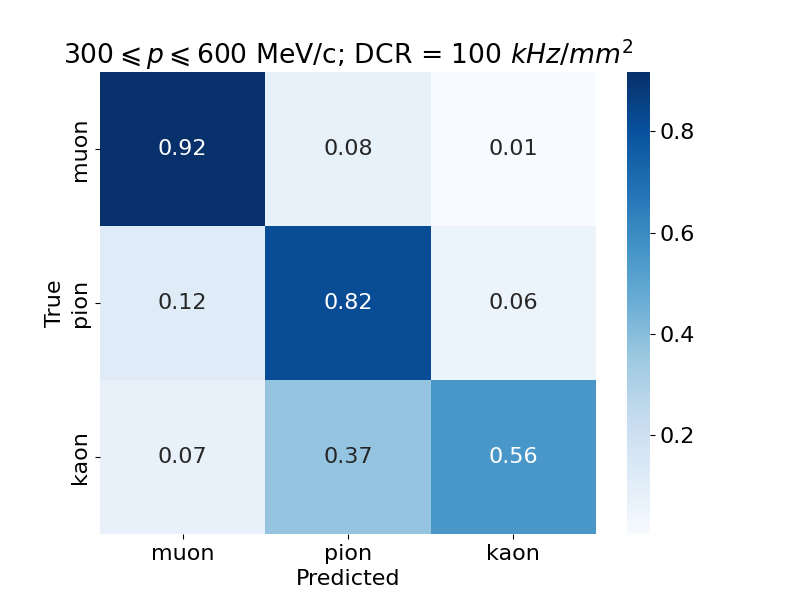}
    \includegraphics[width=.49\textwidth]{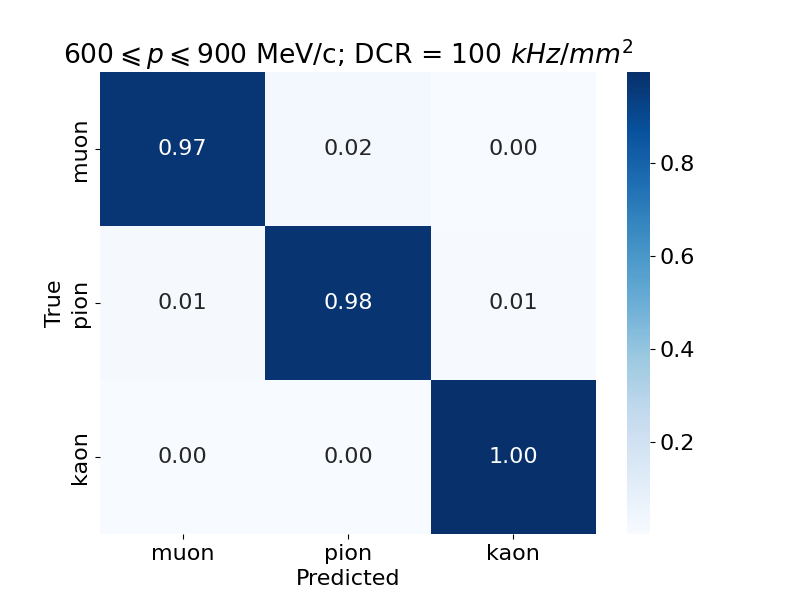} \\
    \includegraphics[width=.49\textwidth]{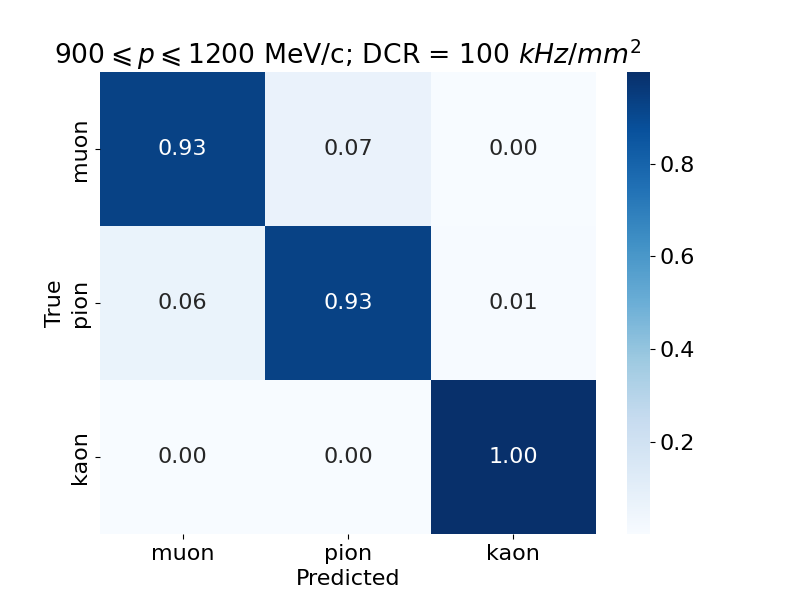} 
    \includegraphics[width=.49\textwidth]{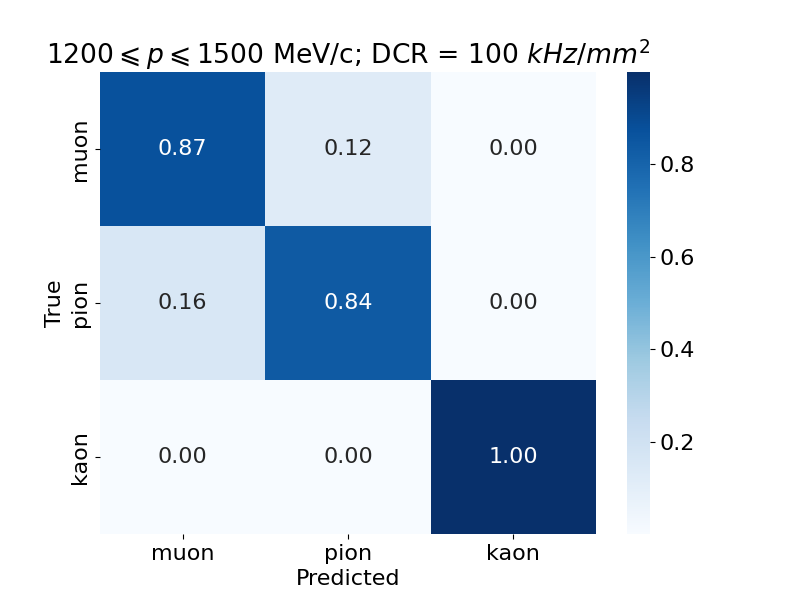} \\
    \includegraphics[width=.49\textwidth]{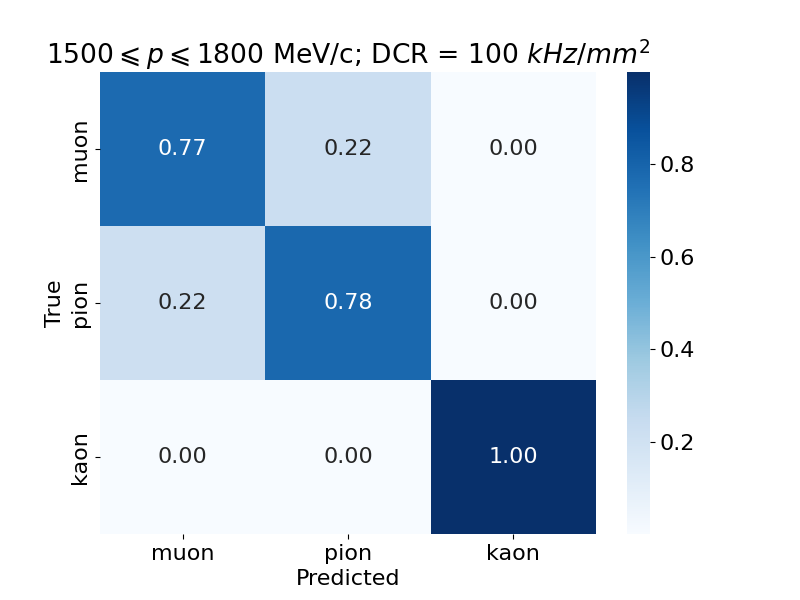} 
    \includegraphics[width=.49\textwidth]{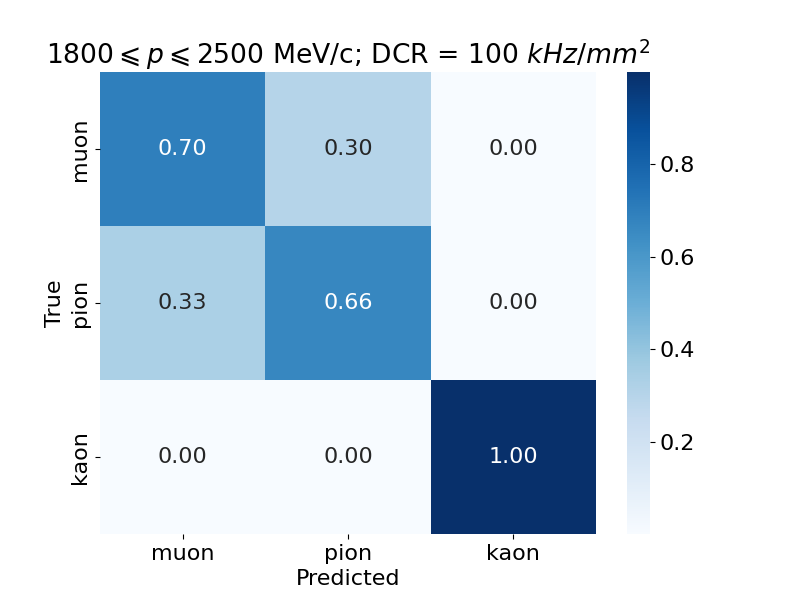} 
    \caption{\label{fig:conf_mat_dcr5} Confusion matrices (fraction of predicted type from true generated type) for the test sample of single particles for $\mathrm{DCR} = 100\khzmm$  in six momentum ranges.}
\end{figure}

\begin{figure}[h]
    \includegraphics[width=.49\textwidth]{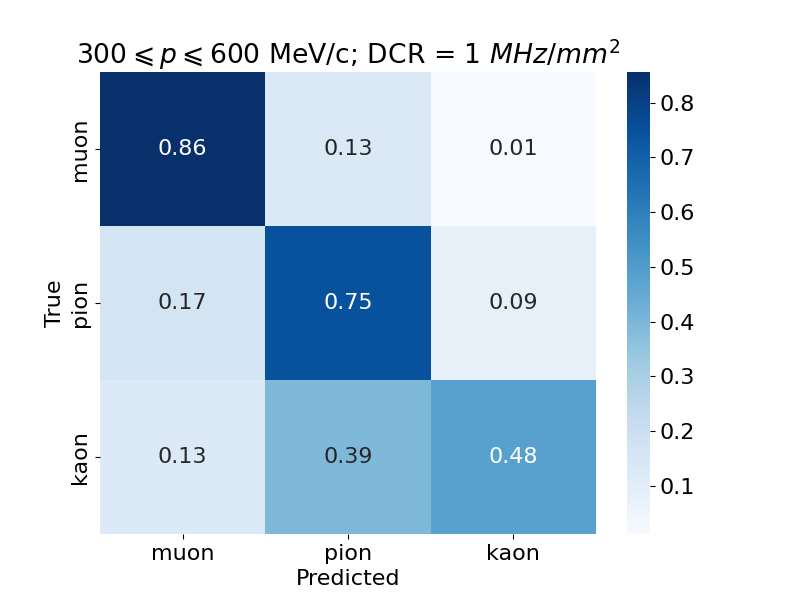}
    \includegraphics[width=.49\textwidth]{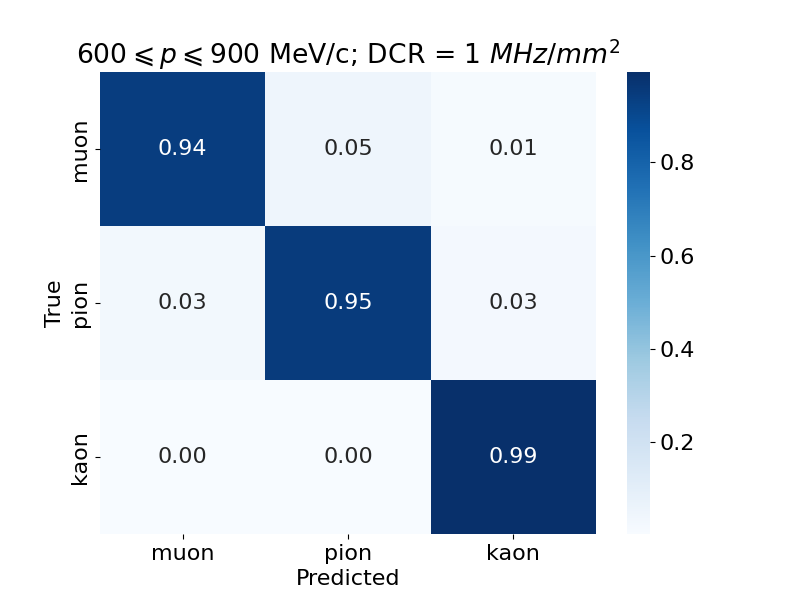} \\
    \includegraphics[width=.49\textwidth]{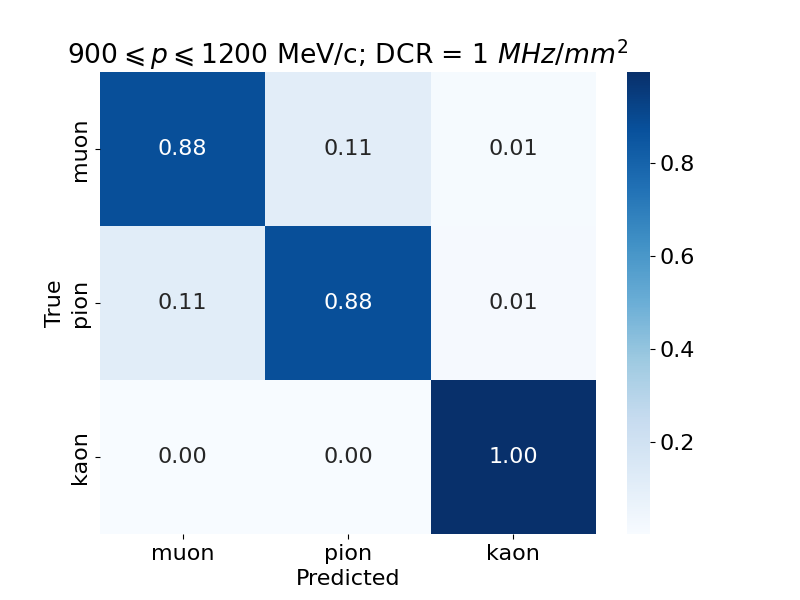} 
    \includegraphics[width=.49\textwidth]{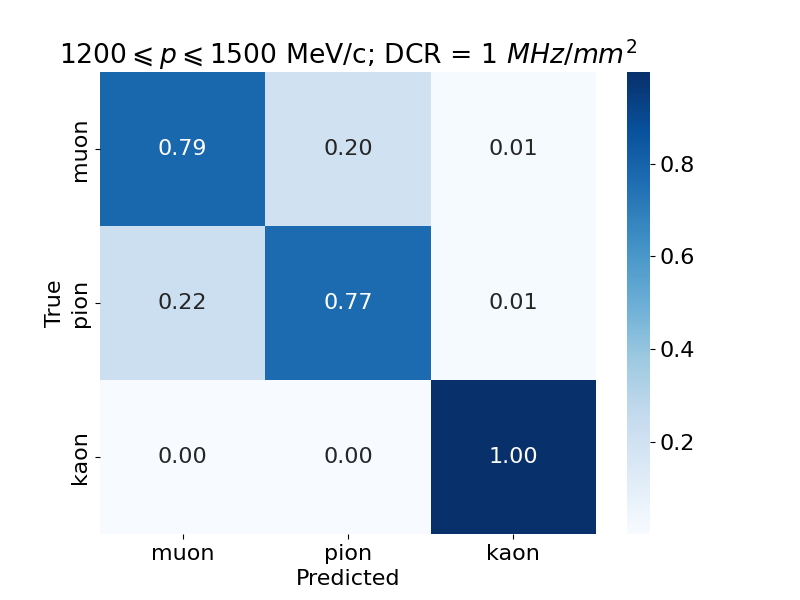} \\
    \includegraphics[width=.49\textwidth]{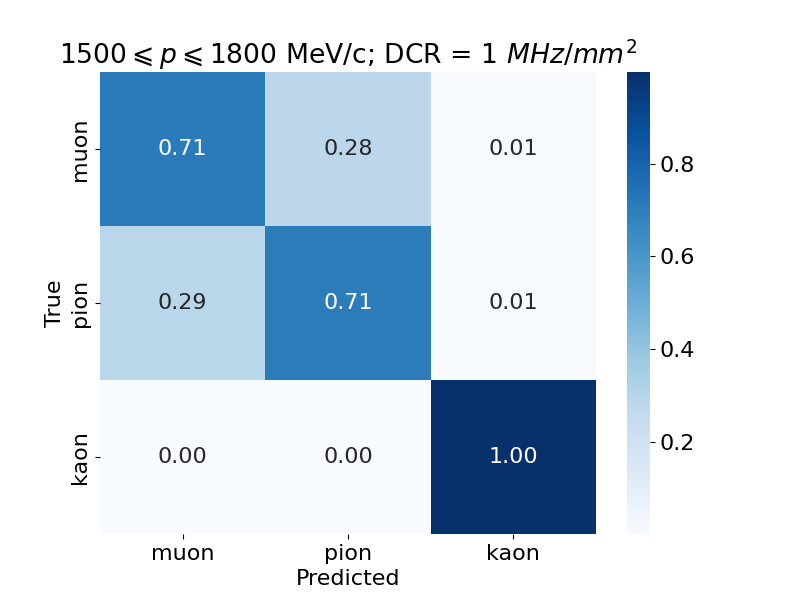} 
    \includegraphics[width=.49\textwidth]{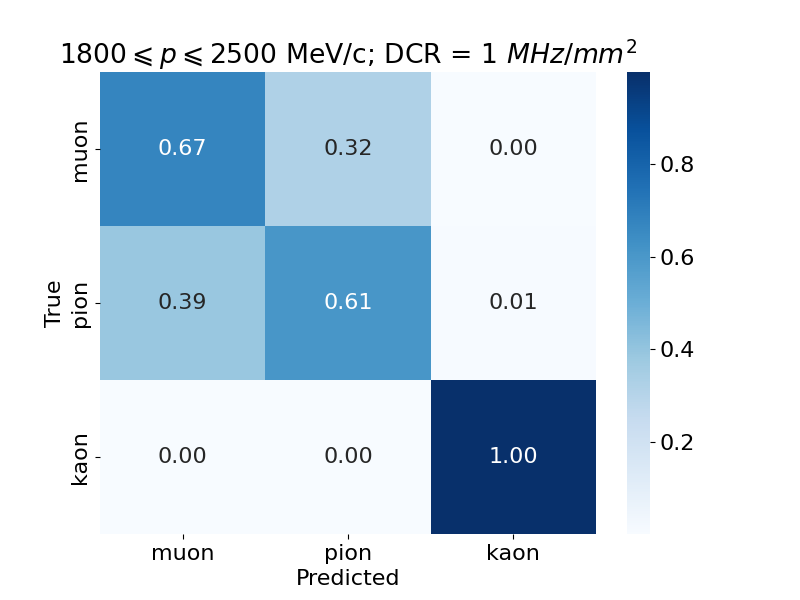} 
    \caption{\label{fig:conf_mat_dcr6} Confusion matrices (fraction of predicted type from true generated type) for the test sample of single particles for $\mathrm{DCR} = 1\mhzmm$ in six momentum ranges.}
\end{figure}

The single BDT classifier is trained for the whole momentum range. However, the PID performance differs significantly with momentum. Therefore, the confusion matrices for the test sample are presented  in six momentum ranges in figures~\ref{fig:conf_mat_dcr5} and \ref{fig:conf_mat_dcr6} for two DCR values. At low momenta reconstruction efficiency suffers from the low number of detected Cherenkov photons, which affects identification quality. As a result, the first momentum bin with 300~$ \le p \le$~600~$\mev/c$ is the only one where kaon misidentification has a noticeable probability.

The SiPM geometrical resolution makes it impossible to distinguish between muons and pions with momenta above $\sim1800~\mevc$, which leads to an inevitable drop in classification quality in this range with FARICH as the only PID  subsystem. For this reason the last bin covers the range $p \ge 1800~\mevc$. Information on the number of signal hits in each event, which is incorporated into the z-score, allows a certain level of identification above the $1800~\mevc$ threshold, though.

\begin{figure}[h]
    \includegraphics[width=.49\textwidth]{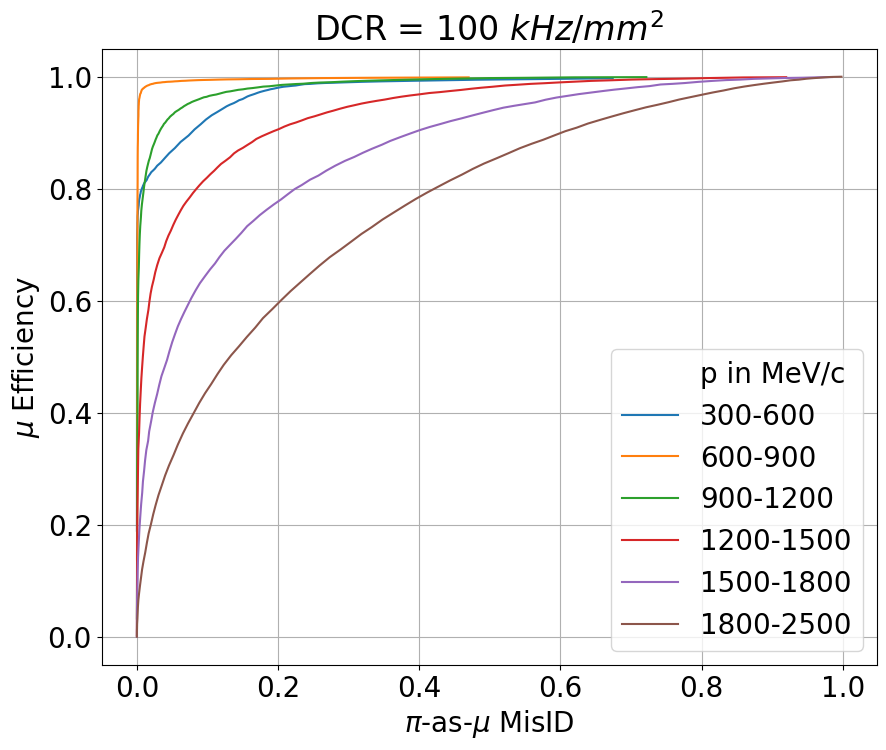}
    \includegraphics[width=.49\textwidth]{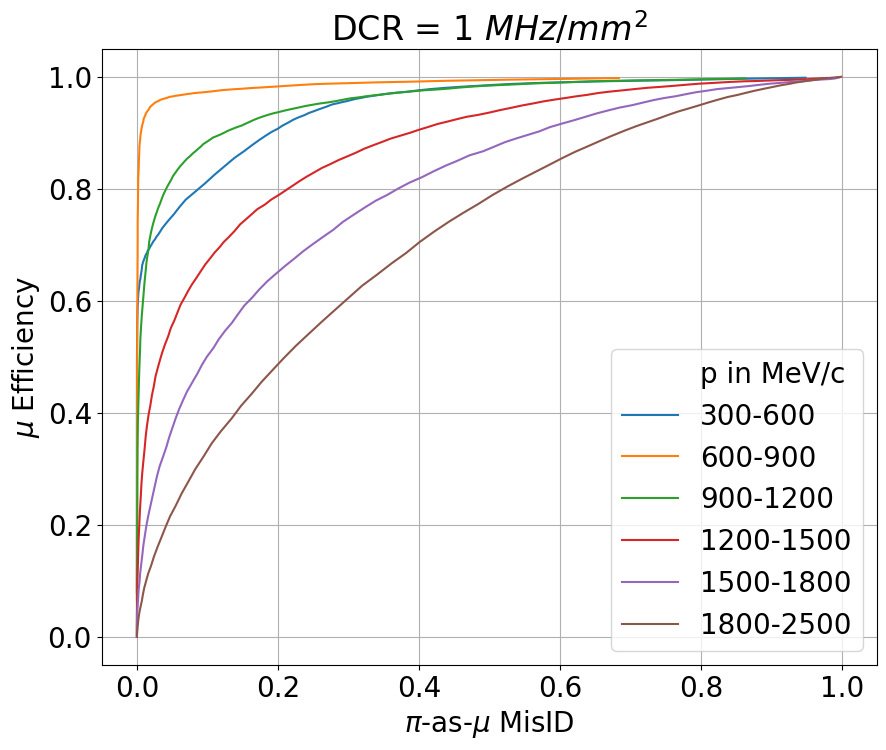}
    \caption{\label{fig:eff-misid_single} Muon identification efficiency versus pion-as-muon misidentification probability for the test sample of single particles for $\mathrm{DCR} = 100\khzmm$ (left) and $\mathrm{DCR} = 1\mhzmm$ (right) in six momentum ranges.}
\end{figure}

The muon identification efficiency and misidentification probability of pion as muon are shown for the test sample in figure~\ref{fig:eff-misid_single} for two DCR levels in several momentum bins. For the DCR level of $1\mhzmm$ the drop of identification performance can be seen in all momentum ranges. The PID drop in the momentum range below $600\,\mevc$ is related to significant confusion of both pions and muons with kaons. 

\section{Projection of PID performance to physics analysis}
\label{sec:performance}
The analysis of measurement precision of branching fraction of $D^0\rightarrow K\mu\nu$ decay has been chosen to demonstrate the PID performance. It should be noted that the full reconstruction at the model of \sctf detector is not available yet and the estimates are obtained for different assumptions about tracker and calorimeter performances.

The decays of $\psi(3770)$ resonance, which is an admixture of $2S$ and $1D$ $c\bar{c}$ states, can be considered as a natural source of $D$~meson pairs at \ee collider experiments. The state $\psi(3770)$ decays into $D^0\bar{D}^0$ pair with the probability of $52^{+4}_{-2}\%$~\cite{ParticleDataGroup:2024cfk} without any extra final state particles. With the design peak instantaneous luminosity about 10$^{35}$~cm$^{-2}$s$^{-1}$, the annual integrated \stcf luminosity at $\psi(3770)$ peak reaches $\sim$150~fb$^{-1}$, which corresponds to $\mathcal{O}(10^9)$  of $\psi(3770)$ produced~\cite{ref:Achasov:2024}. Taking advantage of this enormous data sample, the experiment at the \stcf can test LFU with unprecedented precision. As was already mentioned above, the promising channels for this task are \eg $D^0\rightarrow K\mu\nu$ and $D^0\rightarrow Ke\nu$. While the efficiency of the latter channel relies on the electromagnetic calorimeter, the reconstruction of the channel $D^0\rightarrow K\mu\nu$ significantly depends on the particle identification performance.

The estimates made in this study follow the strategy developed by BESIII collaboration \cite{ref:besIII_2019}, which implies event tagging by reconstructing one of the D mesons, while from the other only kaon and muon are reconstructed. As a conservative estimate, we use the efficiency of the decay chain reconstruction averaged over three tag-side channels\footnote{Charge conjugation is implied through the paper.}: $\bar{D}^0_\mathrm{tag}\to K^+\pi^-,\, K^+\pi^-\pi^-\pi^+,\,K^+\pi^-\pi^0$. The efficiency and total number of the tagged events taken from \cite{ref:besIII_2019} are $\varepsilon=58.93\%$ and $N_\mathrm{tag}^\mathrm{BESIII}=2341408$, respectively. The number of the tags is then scaled with expected integrated luminosity by factor of $\frac{\mathcal{L}^{\sctf}}{\mathcal{L}^{BES III}}=\frac{150~fb^{-1}}{2.93~fb^{-1}}$ yielding $N_\mathrm{tag}^\mathrm{\sctf}=7.99\cdot10^8$ tagged $D^0$. The main background comes from the $D^0\to K^-\pi^+\pi^0$ decay, which could mimic the signal in case of lost $\pi^{0}$ and $\pi^+/\mu^+$ misidentification. To suppress possible background from the $D^0\to K^- \pi^+$ decays the invariant mass of the $K^-\mu^+$ combination will be required to be lower than $1.56\,\gevcc$. 

To estimate the performance of the developed PID algorithm for the physics analysis discussed above, about 550000 events with decays $\psi(3770) \rightarrow D^0\bar{D}^0$ at the centre-of-mass energy of \ee \,beams $\sqrt{s} = $~3770~MeV have been generated and propagated through the \sctf detector. The produced $\bar{D}^0$ is forced to decay into invisible particles and is further ignored. The $D^0$ meson is forced to decay as either $D^0\to K^- \mu^+\nu_\mu$ (signal) or $D^0\to K^- \pi^+ \pi^0$ (background). The propagation of particles in a drift chamber volume in 1~T magnetic field as well as generation of Cherenkov photons in the FARICH subsystem and hits detection are simulated within the Aurora framework. 

For the performance analysis, events are selected using generator level information about the particle momenta. The selection efficiencies listed below affect signal-to-background ratio and do not depend on the PID performance. 
\begin{itemize}
    \item The ratio of number of signal and background events is taken from the corresponding branching fractions from  PDG~\cite{ParticleDataGroup:2024cfk}: $\mathcal{B}_{signal} = \mathcal{B}(D^0\to K^- \mu^+\nu_\mu) = $~0.034 and $\mathcal{B}_{bgd} = \mathcal{B}(D^0\to K^- \pi^+ \pi^0) = $~0.144.
    \item From the background events, only those with lost $\pi^{0}$ are retained. The probability of this case, $P_{lost\pi^{0}}$, is combined from two efficiencies, $\epsilon_{2\gamma}$ and $\epsilon_{\pi^{0}}$. The efficiency $\epsilon_{2\gamma}$ is the probability to detect both $\gamma$s from $\pi^{0}$, i.e. that both $\gamma$s reach a calorimeter (barrel or endcap) and have energies $E_{\gamma} > E_{\gamma cut}$. The geometrical acceptance estimated for the \stcf calorimeter covers the polar angles $|\theta| < $~70.8\textdegree~\cite{ref:Achasov2023} and is fixed for this study, it amounts to 0.95 for single $\gamma$. The efficiency $\epsilon_{\pi^{0}}$ is the probability to reconstruct  $\pi^{0}$ from two $\gamma$s, which is related to the chosen value of $E_{\gamma cut}$. The probability to reconstruct $\pi^{0}$ depends on the calorimeter performance, which is not known yet. For this reason, several combinations of  $\epsilon_{\pi^{0}}$ and $E_{\gamma cut}$ are considered for analysis.
    \item Both charged particles (tracks) must go through the barrel part of the \sctf detector with the FARICH subsystem. For the fixed detector geometry and very similar kinematics of signal and background, this efficiency is $\epsilon_{tr}^{signal} \approx \epsilon_{tr}^{bgd} \approx $~0.58 for two tracks.
    \item The momenta of both charged particles (tracks) must exceed 310~Mev/$c$, i.e. the limit sensitivity of the FARICH subsystem. The estimated efficiencies are $\epsilon_{F}^{signal} \approx $~0.84 and  $\epsilon_{F}^{bgd} \approx $~0.78.
\end{itemize}

The numbers of signal and background events, ${N_{sel}(K\mu\nu)}$ and $N_{sel}(K\pi\pi^0)$, selected before identification, can be calculated from the efficiencies listed above as 
\begin{equation}
\label{eq:nsignalsel}
    N_{sel}^{signal} = N_{gen,signal} \cdot \epsilon_{tr}^{signal} \cdot \epsilon_{F}^{signal}
\end{equation}
\begin{equation}
\label{eq:nbgdsel}
    N_{sel}^{bgd} = N_{gen,bgd} \cdot (1 - \epsilon_{2\gamma} \cdot\epsilon_{\pi^{0}}) \cdot\epsilon_{tr}^{bgd} \cdot \epsilon_{F}^{bgd}, \quad
    N_{gen,bgd} = N_{gen,signal} \cdot \frac{\mathcal{B}_{bgd}}{\mathcal{B}_{signal}}, 
\end{equation}
\noindent where $N_{gen,signal}$ is the number of generated signal events (about 100000). As can be seen from eq.~\ref{eq:nsignalsel} and \ref{eq:nbgdsel}, in addition to particle identification characteristics, the efficiency of $\pi^{0}$ reconstruction also affects the background admixture to the signal sample under study. The efficiencies of signal and background event reconstruction and numbers of selected events for the four probabilities of lost $\pi^{0}$ are summarised in table~\ref{tab:eff}. 

   \begin{table}[]
    \caption{Branching fractions, reconstruction efficiencies and numbers of selected signal and background events for the BDT performance estimates at various conditions of $\pi^{0}$ reconstruction. }
      \centering
       \begin{tabular}{|c|c|c|c||c|c|c|c|c|}
       \hline
          ${\mathcal{B}}_{signal}$ & $\epsilon_{tr}^{signal}$ & $\epsilon_{F}^{signal}$ & $N_{sel}^{signal}$ & ${\mathcal{B}}_{bgd}$ & $\epsilon_{tr}^{bgd}$ & $\epsilon_{F}^{bgd}$ & $P_{lost\pi^{0}}$ [$\epsilon_{\pi^{0}}$, $E_{\gamma cut}$] & $N_{sel}^{bgd}$ \\
         \hline
        & & & & & & & 0.68 [0.9, 50~MeV] & 65293 \\
          \cline{8-9}
        0.034 & 0.58 & 0.84 & 47415 & 0.144 & 0.58 & 0.78 &  0.60 [0.8, 50~MeV] & 80352 \\
          \cline{8-9}
         & & & & & & & 0.19 [0.9, 200~MeV] & 164219 \\
          \cline{8-9}
         & & & & & & & 0.17 [0.8, 200~MeV] & 168285 \\
         \hline             
       \end{tabular}
       \label{tab:eff}
   \end{table}

\begin{figure}[h]
\begin{center}
        \includegraphics[width=.7\textwidth]{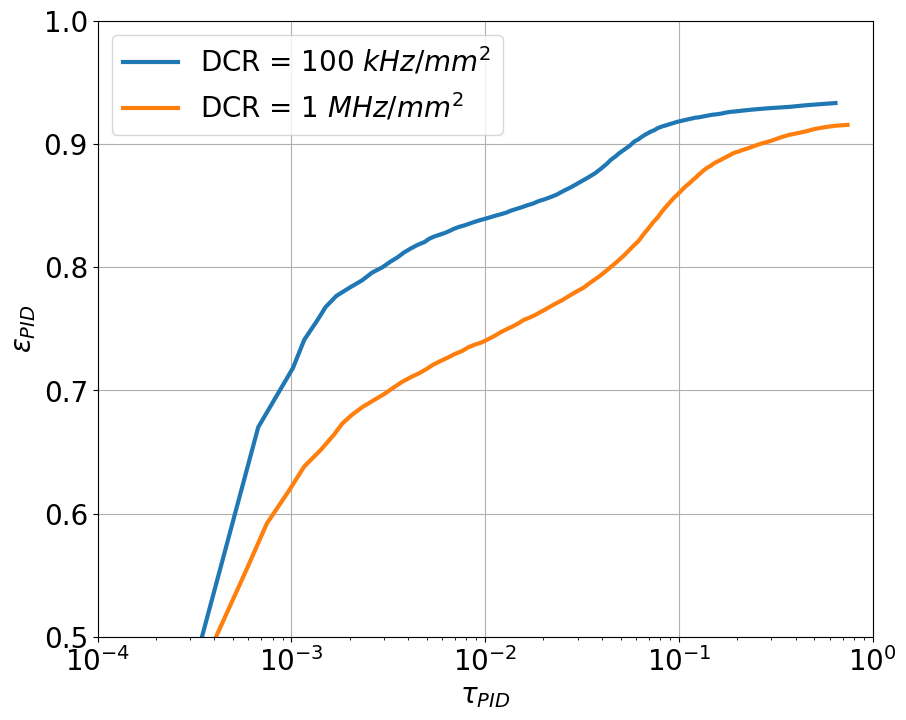}
\end{center}
    \caption{\label{fig:eff_misID_Dmeson} 
    Signal identification efficiency versus background misidentification probability for the expected mixture of signal ($D^0\rightarrow K\mu\nu$) and background ($D^0\rightarrow K\pi\pi^0$) events for $\mathrm{DCR} = 100\khzmm$ (blue) and $\mathrm{DCR} = 1\mhzmm$ (orange). The lines are shown to guide the eye. See text for details.}
\end{figure}

  The ring reconstruction algorithm described above and the BDT inference to identify a particle specie are applied to selected events and numbers of identified signal and misidentified background events are calculated. The efficiency of kaon identification, $\epsilon_{K}$, is taken into account: negatively charged tracks are required to be identified as kaons. The numbers of selected and reconstructed signal events identified as signal can be represented as follows:
  \begin{equation}
  \label{eq:nsignal}
      N_{reco}^{signal}(K\mu) = N_{sel}^{signal} \cdot \epsilon_{PID}; \quad \epsilon_{PID} = \epsilon_{K} \cdot \epsilon_{\mu},
  \end{equation}
  \noindent where $\epsilon_{\mu}$ is the efficiency of muon identification in the FARICH subsystem. In turn, the number of background events misidentified as signal can be calculated as follows:
  \begin{equation}
  \label{eq:nbgd}
      N_{reco}^{bgd}(K\mu) = N_{sel}^{bgd} \cdot \tau_{PID}; \quad \tau_{PID} = \epsilon_{K} \cdot \tau_{\pi/\mu},
  \end{equation}
   \noindent where $\tau_{\pi/\mu}$ is the misidentification probability of pion as muon. From eq.~\ref{eq:nsignal} and \ref{eq:nbgd}, the pairs of $\epsilon_{PID}$ and $\tau_{PID}$ values can be obtained for various BDT working points defined by the classifier threshold at different assumptions about the probabilities of lost $\pi^{0}$. The efficiency of signal identification $\epsilon_{PID}$ versus the probability of background misidentification $\tau_{PID}$ is shown in figure~\ref{fig:eff_misID_Dmeson} for two noise levels and averaged over four assumptions about $\epsilon_{2\gamma}$ and $\epsilon_{\pi^{0}}$ from table~\ref{tab:eff}, which result in very similar BDT performance. 

\begin{figure}[h]
\begin{center}
     \includegraphics[width=.7\textwidth]{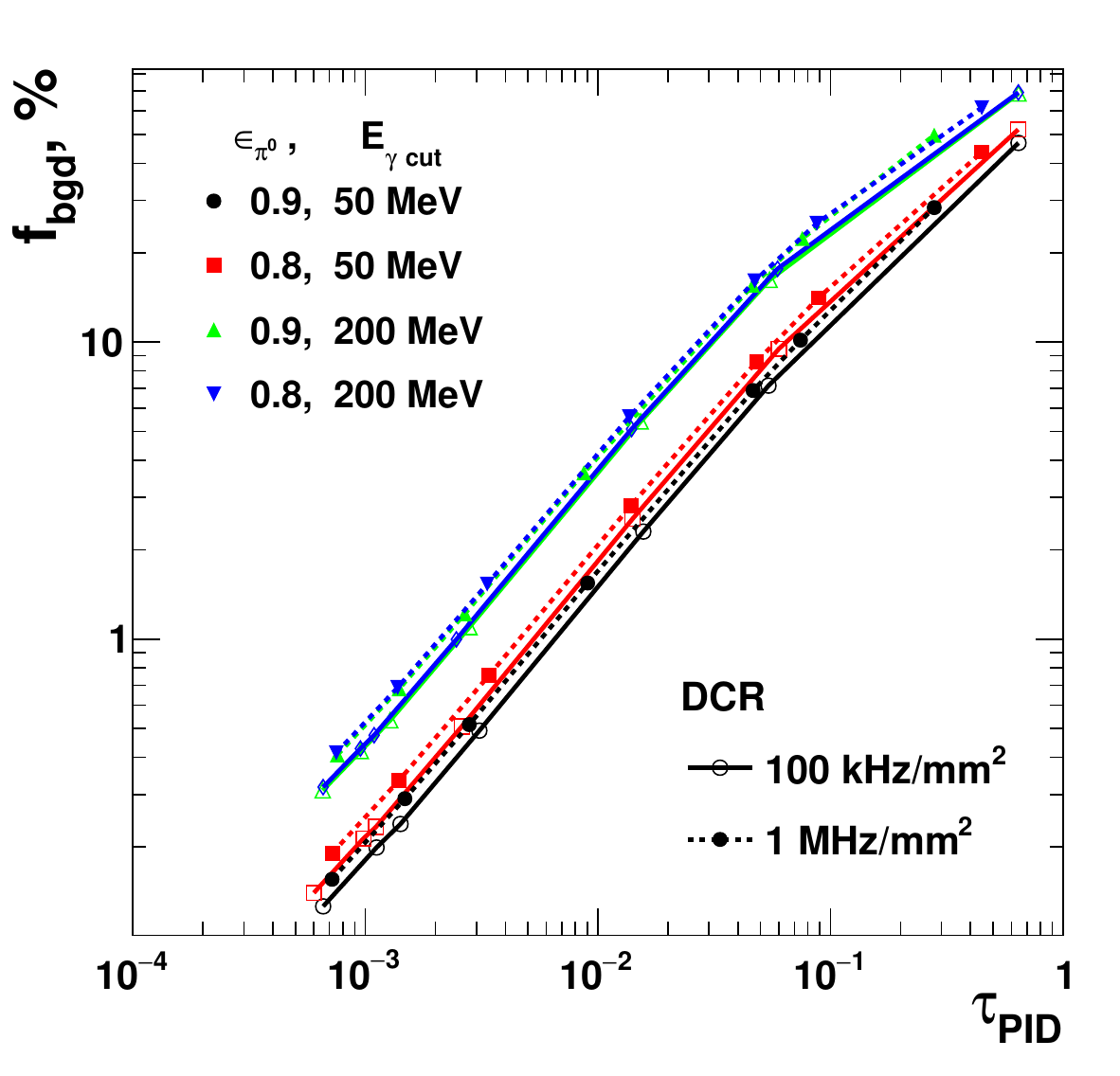}   
\end{center}
    \caption{\label{fig:bgd_misid} Expected background fraction in the reconstructed signal sample versus the background misidentification probability for different efficiencies of $\pi^0$ reconstruction at $\mathrm{DCR}= 100\khzmm$  (solid lines) and $\mathrm{DCR} = 1\mhzmm$ (dashed lines). The lines are shown to guide the eye.}
\end{figure}

 The unprecedented expected luminosity of charm superfactories together with the achieved performance of particle identification lead to the negligible statistical uncertainty for the $D^0\to K\mu\nu$ branching ratio measurements at the level well below 0.1\%.  Figure~\ref{fig:bgd_misid} shows the percentage of background contribution, $f_{bgd}$, for various assumptions about $\pi^0$ reconstruction efficiencies and conditions and for two noise levels. With the possibility to move the background admixture below 0.5\%, the systematic uncertainty can be also made smaller opening the opportunity to achieve a new level of LFU measurement precision at charm superfactories.

\section{Conclusion}
\label{sec:conclusion}

The performance of the FARICH-based PID system for the future charm superfactory is investigated using simulated samples of single particles as well as simulation of D meson decay processes. The algorithms of Cherenkov rings reconstruction and BDT-based particle identification have been developed and tested for different event reconstruction conditions. The algorithm robustness  under various SiPM dark count rates was proved: it was shown, that $\pi/\mu$ separation at the level of 5 standard deviations is available up to $\sim 700~\mevc$ for the most conservative $\mathrm{DCR} = 1\mhzmm$ and up to $\sim 1.3~\gevc$ for the baseline $\mathrm{DCR} = 100\khzmm$. 

With the physics analysis of the $D^0\to K\mu\nu$ decay, the FARICH PID performance is further explored. 
Taking into account the most conservative assumptions about the subsystem efficiencies and background admixture due to misidentification as well as the planned luminosity of the charm superfactory, the role of the PID efficiency provided by a single subsystem of the \sctf is demonstrated.  
 For this physics case, the dominated contribution to background rejection comes from the particle identification with the FARICH subsystem, even for the conservative expectations of the photosensor dark count rates. 
As follows from the analysis, the performance of the developed algorithms allows significant mitigation of the background contribution and therefore improvement of the measurement precision.

\section*{Acknowledgments}

The article was prepared with partial support of the project
"Mirror Laboratories" of the  HSE University.

\bibliographystyle{JHEP}
\bibliography{biblio.bib}

\end{document}